# Copula-Based Univariate Time Series Structural Shift Identification Test

Henry Penikas

An approach is proposed to determine structural shift in time-series assuming non-linear dependence of lagged values of dependent variable. Copulas are used to model non-linear dependence of time series components. Several nice properties of copula application to time series are discussed. To identify the break copula structural shift test is applied. Data on quarterly GDP growth rate for the US from 1947 till 2012 is used as an empirical example. It is shown that the proposed approach captures the recession of 1981-1982 as the key break date in GDP growth rate series time structure that cannot be identified by standard time series structural break tests.

**JEL Codes**:
*C46* [Modeling specific distributions],
*C14* [Semiparametric and Nonparametric Methods: General].

*Henry Penikas* - Lecturer at Mathematical Economics and Econometrics Department, Senior Research Fellow at International Laboratory of Decision Choice and Analysis, National Research University Higher School of Economics. E-mail: penikas@gmail.com

# 1. Introduction

The issue of structural breaks in time series was originally investigated by Wald (1947). The problem has attracted a lot of attention since; see Perron (2005) for a comprehensive review. Recently the area of joint distributions has attracted significant importance given applications in financial risk management and actuarial science. Copula models are often used in this context when it is useful to decompose joint distribution modeling into two steps: modeling marginal distributions and dependence (i.e. copulas themselves). As a matter of fact structural tests for copulas started being developed to research the stability of copulas during the time. Among the recent papers one might find Harvey (2008), Brodsky et al. (2009), Patton (2012), Holmes et al. (2013), Quessy et al. (2013).

The objective of the current paper is to research univariate, not multivariate time series using same copula structural shift identification approach. The paper contributes in several ways:

- The unique feature of time series in terms of copula is presented. Particularly, the stability of marginals for various lagged components of the series;
- Conventional (in copula theory) independence test is interpreted as a non-linear correlogram test;
- Copula structural break test procedure application enables to reveal shifts, missed by conventional (linear) tests on structural break identification (empirical example of US GDP is considered).

As a result the paper is organized as follows. First Section 2 is devoted to brief literature review. Second theoretical framework is given in Section 3. Then Section 4 presents the data used. Section 5 provides the results of test procedure application. Section 6 concludes.

# 2. Literature Review

The most common structural break tests are that of Andrews-Zivot (e.g. Andrews (1993)) and Philips-Perron (cf. Perron (2005)). The idea is to consider the change in intercept and (or) trend for the linear time series model. A dummy variable approach is used to detect the moment when the change is significant to be considered as the break point.

Previous works dealing with copula structural break identification of similarly copulas comparison included Genest, Remillard (2004); Remillard, Scaillet (2006); Tsukahara (2007); Brodsky et al. (2009).

Before discussing copula models application to time series analysis it is necessary to be fair enough and mention the works of Darsow et al. (1992) and Ibragimov (2009) who already researched the properties of copulas when applied to Markov processes. Particularly, Ibragimov (2009) defines r- and m-dependence properties for copulas to be suitable for time series modeling.

# 3. Theoretical Framework

Copulas represent a way of joint probability distribution function decomposition as it is given below in (1). Extensive overview of copulas and their properties as the linkage to triangular norms might be found in Nelsen (2006) and Alsina at al. (2006), respectively.

$$H(x_1,...,x_d) = C(F_{X_1}(x_1),...,F_{X_d}(x_d)) \quad (1)$$

$$H(y_t,...,y_{t-d+1}) = C(F_Y(y_t),...,F_Y(y_{t-d+1})) \quad (2)$$

*Stationarity Hypothesis 1. Marginal distributions when decomposing time series into copula and marginal are the same.*

$$F_{Y_t} = F_{Y_{t-1}} = ... = F_{Y_{t-d+1}} = F_Y \quad (3)$$

It is necessary to state that in case $X_1 = Y_t$ and $X_d = Y_{t-d+1}$, the following representation (2) holds given (3) that is true for large rows. In case of few observations test restrictions should be studied in greater detail.

The property (3) is of great importance for the testing procedure as it clearly states that having once modeled the marginals their relationship is fully captured by copulas that do not limit the dependence nature to linearity.

Briefly to remind the testing procedure taken from Brodsky et al. (2009).

Two empirical copulas (4) before and after potential break point $l$ are estimated.

$$D_l(u) = \frac{1}{l}\sum_{i=1}^{l} I(U_{i,l} \leq u) = \frac{1}{l}\sum_{i=1}^{l}\prod_{j=1}^{d} I(U_{ij,l} \leq u_j)$$

$$D_{N-l}(u) = \frac{1}{N-l}\sum_{i=l+1}^{N} I(U_{i,N-l} \leq u) = \frac{1}{N-l}\sum_{i=l+1}^{N}\prod_{j=1}^{d} I(U_{ij,N-l} \leq u_j), \quad (4)$$

where $U_{i,l} = (U_{i1,l},...,U_{id,l})$ and for every $j=[1,...,d]$

$$U_{ij,l} = \frac{l}{l+1}F_{j,l}(X_{ij}) = rank(X_{ij})/(l+1), \quad 1 \leq i \leq l,$$

$$U_{ij,N-l} = rank(X_{ij})/(N-l+1), \quad l+1 \leq i \leq N. \quad (5)$$

$N$ is fixed and the following modification of the Kolmogorov-Smirnov statistics (6) is applied:

$$\Psi_{l,N-l}(u) = (D_l(u) - D_{N-l}(u))\sqrt{l(N-l)}/N. \quad (6)$$

Then the statistics (7) takes its maximum value in the break point (8)/

$$T_N = \max_{[\beta N] \leq l \leq [(1-\beta)N]} \sup_u |\Psi_{l,N-l}(u)|. \quad (7)$$

$$\hat{m}_N \in \arg\max_{[\beta N] \leq l \leq [(1-\beta)N]} \left(\sup_u |\Psi_{l,N-l}(u)|\right), \quad (8)$$

Further properties of the test statistics might be found in Brodsky et al. (2009).

As opposed to Brodsky et al. (2009) who considered multivariate time series, the present paper focused on copulas applied not to a time series vector, but to a univariate time series with special attention to the dependence structure of lagged components in it. US GDP official quarterly data is taken as an example. Data description and test results follow below.

## 4. Data Used

To apply non-linear structural break test a very common data row was chosen, i.e. US GDP ranging from 1947 to 2012 sourced from the US Bureau of Economic Analysis. Level and quarterly growth rate data is presented at figure 1 below. The total number of observations equaled to 261.

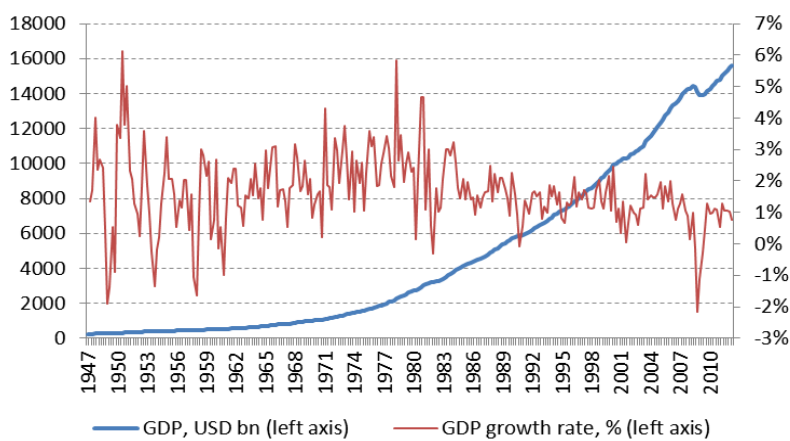

**Figure 1. US GDP Dynamics.**
Source: Bureau of Economic Analysis .

Source: Bureau of Economic Analysis. (URL: http://www.bea.gov/iTable/

iTable.cfm?ReqID=9&step=1)

It is important to note is that GDP time series is rather a low frequency and relatively nonvolatile time series compared to minute- or transaction based financial time series. The latter are prime candidates to assume nonlinear dependence. Nevertheless, it was desirable to start from ordinary low frequency macroeconomic time series for test validation.

Financial time series research might well be subject of another paper where conditional heteroscedasticity might also need revision with respect to non-linearity of variance dependence on its previous values and previous squared residuals' values.

GDP level-data is non-stationary is it follows solely from the visual analysis of data. That is why further analysis was proceeded with the data transformed to growth rates.

Table 1 and Figure 2 below present marginal descriptives proving marginals do not tend to differ given the discrete data. Some deviations in means only exist. This supports the hypothesis made in Section 3 that the successive observations are identically distributed.

**Table 1. Marginals' Descriptives.**

| Lag | 0 | -1 | -2 | -3 | -4 | -5 | -6 | -7 | -8 | -9 |
|---|---|---|---|---|---|---|---|---|---|---|
| Min. | -0,022 | -0,022 | -0,022 | -0,022 | -0,022 | -0,022 | -0,022 | -0,022 | -0,022 | -0,022 |
| 1st Qu. | 0,011 | 0,011 | 0,011 | 0,011 | 0,011 | 0,011 | 0,011 | 0,011 | 0,011 | 0,011 |
| Median | 0,016 | 0,016 | 0,016 | 0,016 | 0,016 | 0,016 | 0,016 | 0,016 | 0,016 | 0,016 |
| Mean | 0,016 | 0,016 | 0,016 | 0,016 | 0,016 | 0,016 | 0,016 | 0,016 | 0,016 | 0,016 |
| 3rd Qu. | 0,021 | 0,021 | 0,021 | 0,021 | 0,021 | 0,022 | 0,022 | 0,022 | 0,022 | 0,022 |
| Max. | 0,061 | 0,061 | 0,061 | 0,061 | 0,061 | 0,061 | 0,061 | 0,061 | 0,061 | 0,061 |

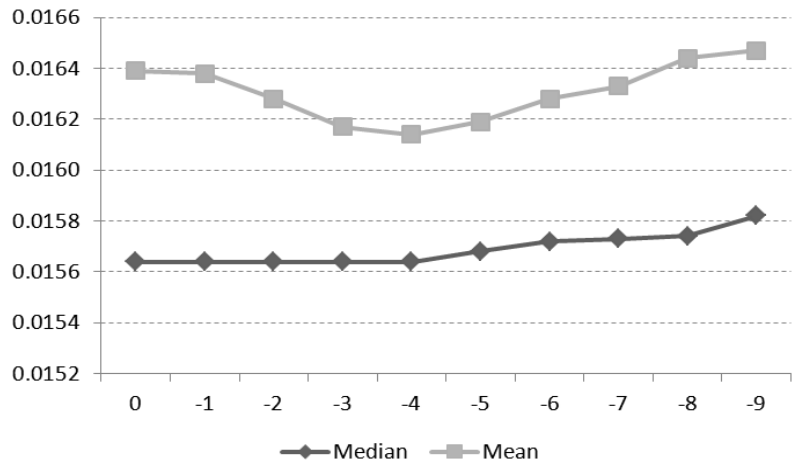

**Figure 2. Mean and Median Values for Marginals.**

Another way of data representation is the scatter plot of current GDP growth rate versus one of its lags. Example for first lag against current values is presented in figure 3. Left part (a) presents actual growth rates, whereas right part (b) shows the respective values of empirical distribution function. The latter graph brings the first hints needed for copula modeling. Though data is not rich in points like highly frequent financial time series, it still some more dispersed values in upper right corner of figure 3 (b) than that of lower left one.

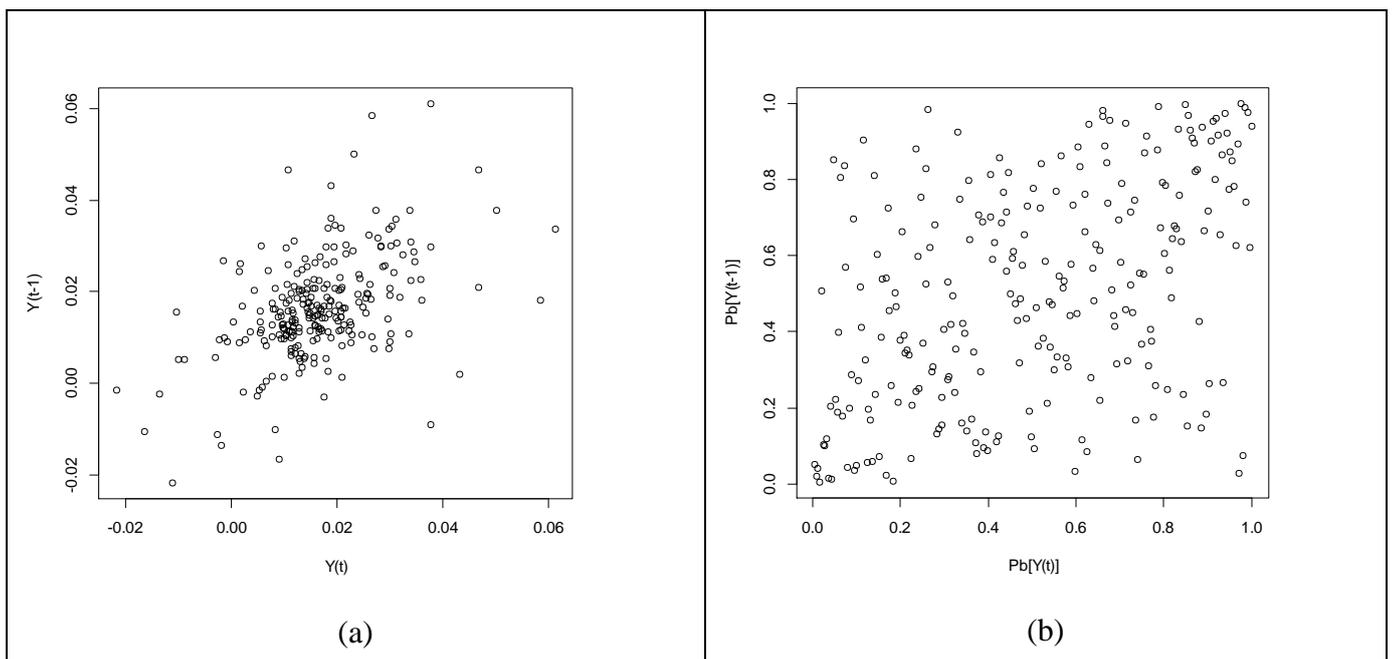

**Figure 3.** Joint distribution of current GDP growth rates and its first lagged values.

# 5. Test Realization

Testing procedure[1] was twofold:

(1) Copula independence test suggested by Genest, Remillard (2004);

(2) Copula structural shift identification as proposed by Brodsky et al. (2009);

When dealing with time series analysis, traditionally one is supposed to look at the correlogram of the row to get the insights on its probable profile in terms of AR and MA components. Figure 4 illustres correlogram for US GDP growth rate series indicating the probable strong dependence of first and second lags to the current value, some jump in ACF is also observed for lags 9 and 10.

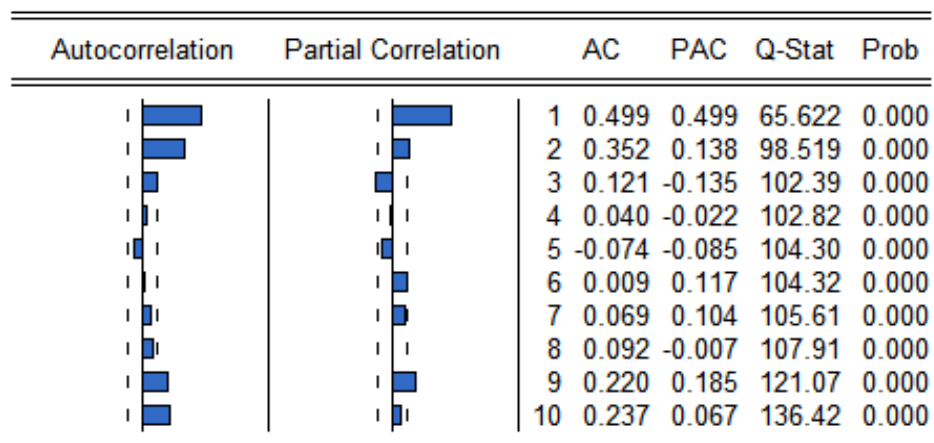

**Figure 4. Correlograms for GDP growth rate series.**

Using a correlogram, it is possible to get an idea as to the probable (statistically significant) correlations between current values and lagged ones. But as we pointed out earlier when interested in non-linear dependence between current and lagged values, one is in need of the test whether the copula joining two values is a product copula corresponding to independence case or not.

It is exactly the question that is answered by Genest and Remillard (2004) Copula Independence Test. The test idea is to compare empirical copula to the product copula. In case the former is not statistically different from the latter, inference about independence of random variables is made. The visual test representation is Dependogram that is given in Figure 5 (reduced case) and in Annex 1 (extensive case). To comment on Figure 5 lines present the test statistics values (they are duplicated in Table 2 below for convenience), dots stand for critical values.

From the non-linear perspective (cf. Annex 1) one can also conclude about statistically significant dependence (or non-independence) case for $1^{st}$, $2^{nd}$, $9^{th}$ lag that is in line with correlogram analysis. The key difference would come when searching for break point based on linearity and non-linearity assumptions.

---

[1] Copula independence test and copula structural break test were run in R software, whereas Andrews-Zivot test was done in EViews environment.

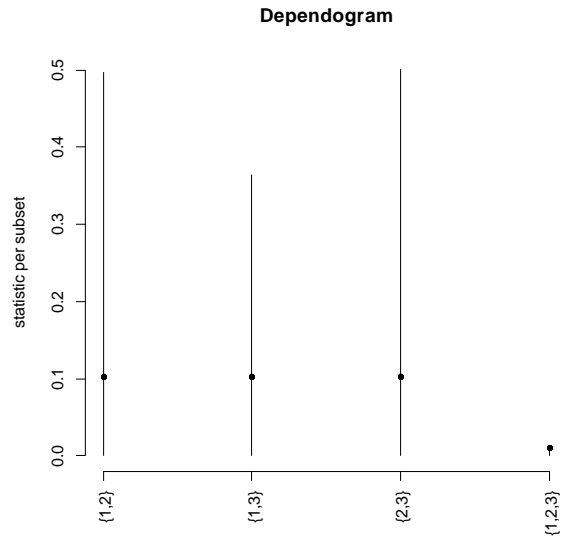

**Figure 5. Selected Copula Dependogram (for 1 and 2 lags)**

Different to table in Annex 1 that gives one idea on bivariate copulas (i.e. pairwise dependence), Figure 5 above presents the case of a trivariate dependence, e.g. case {1,2,3}. Trivariate case analysis tests where current value of US GDP growth rate and its 1st and 2nd lag together are dependent or not. As Table 2 below shows the respective test statistics exceeds the critical value though not substantially. This implies the necessity to proceed with bivariate copula analysis with respect to time series. Never-the-less, it does not exclude application of hierarchical (cf. Okhrin et al. (2009)) or vine- (cf. Cooke et al. (2011)) copulas.

**Table 2. Copula Independence Test Statistics Values.**

|   | subset | statistic | pvalue | critvalue |
|---|--------|-----------|--------|-----------|
| 1 | {1,2}  | 0.497502  | 0.00495 | 0.103565 |
| 2 | {1,3}  | 0.363827  | 0.00495 | 0.103565 |
| 3 | {2,3}  | 0.501566  | 0.00495 | 0.103565 |
| 4 | {1,2,3}| 0.009405  | 0.05445 | 0.010610 |

Our second step is to directly apply the copula structural break test to searching the break point in the time series. Figure 6 below presents the test statistics dynamics. To remember as shown in Section 3 and proven in Brodsky et al. (2009) statistics maximum corresponds to the break point. Here the bivariate[2] case of $H(y_t, y_{t-1}) = C(F_Y(y_t), F_Y(y_{t-1}))$ is considered. The break point as summarized in table 3 below is 4th quarter of 1981.

---

[2] Testing up to 10-copula structural break resulted in a similar date of 1981.

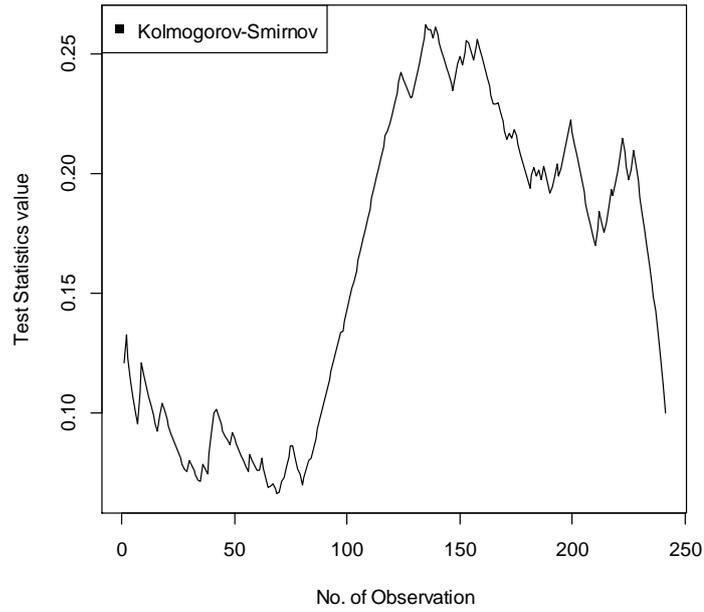

**Figure 6.** Copula Structural Break Statistics for US GDP Growth Rate Time Series.

To give one an idea what has changed in terms of copula with the dependence for US GDP growth rate and its 1$^{st}$ lag Figure 7 is presented. Left part (a) of Figure 7 is more dispersed with the presence of points concentration in the right upper corner implying that Gaussian or Gumbel copulas might better descibe them. The latter is characterized by non-zero dependence of upper tails of distribution.

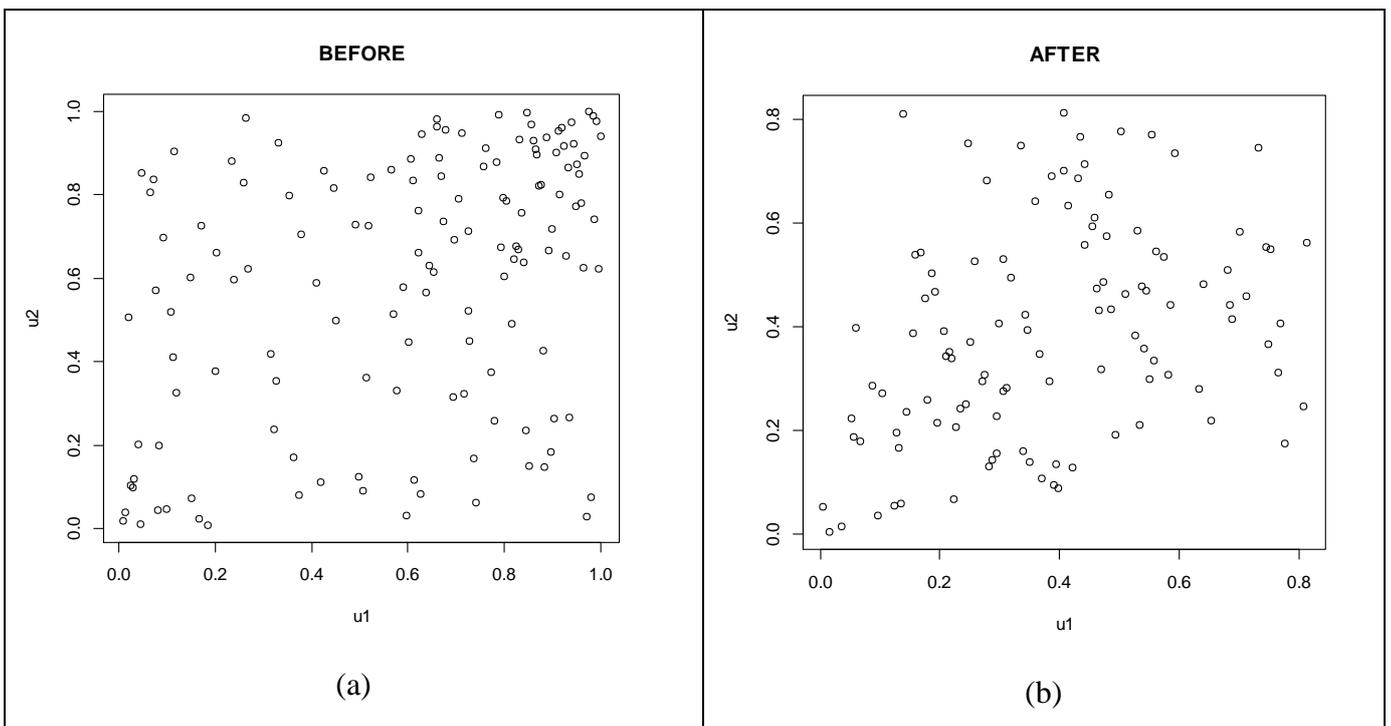

**Figure 7.** Scatterplots Corresponding to Copulas Before (a) and After (b) the Identified Break Date.

Inversely right part (b) of Figure 7 is less dispersed and more concentrated to lower left corner suggesting Clayton copula might be the best candidate to describe such a dependence profice. To mention Clayton copula is characterized by the non-zero dependence of lower tails of distribution (to remind Gaussian copula has zero tail dependence for both upper and lower tails of distribution).

The proposed approach is afterwards benchmarked to the conventional Andrews-Zivot test output as one may see in Table 3. All there test versions of Andrews-Zivot test were checked, including test for structural break in intercept, trend or in both.

**Table 3. Comparison of Structural Break Tests.**

| Test Type | Test Specification | Shift Observation | Shift Date | Test Statistics |
|---|---|---|---|---|
| Copula | Kolmogorov-Smirnov | 140 | 1981Q4 | 0.2622 |
| Andrew-Zivot | Intercept | 72 | 1964Q4 | -10.3745 |
| | Trend | 123 | 1977Q3 | -10.3023 |
| | Intercept + Trend | 96 | 1970Q4 | -10.5476 |

Table 3 evidently shows that Andrews-Zivot linear test brings us the result of changes happening in 1964, 1970, 1977 (Annex 2 provides details for the output of testing procedure for the break point). Perhaps this is the reason for Perron to state that the problem of structural break detection and understanding is tied to the fact that the break often happens with the delay to the economic root of the structural break.

Nevertheless, it is interesting to refer to the guidance on the US history to trace what facts might underline the change. [TFC] materials suggest that the United States of America faced the first recession after Great Depression, and it was in 1973-1975 linked to the world oil crises. Thus the Andrews-Zivot test including the trend might well illustrate the delayed effect of oil crisis. Still the dates of 1964 and 1970 cannot be that explicitly explained. Not to mention the problem of reverse-engineering that having no external knowledge on economic environment one might fail to choose the correct structural break date between the three: 1964, 1970 and 1977.

What is more interesting, is that observing (and, of course, firstly assuming) non-linear nature of dependence in time series components, end of 1981 is found as the structural break date. When reverting to the US history [TFC], one may recall the events of 1981-1982 when the Iranian Revolution forced oil prices to increase once again. As one can see, the linear test for structural shift identification was unable to detect the date of 1981-1982. The latter date was the last in a sequence of crisis events, as the next recession linked with the Gulf War took place only 10 years later, i.e. in 1990-1991.

# 6. Concluding Remarks

Current paper presented the copula structural shift test application to testing for structural shift in a univariate time series compared to conventional linear testing procedures.

The key findings are as follows:

- A nice hypothesis of time series components is noted, i.e. the equality of marginal distributions. Using copula decomposition, this property enables for copula to incorporate all dependence features (both linear and non-linear ones). Then searching for structural break in copula 14 brings one with more information than solely dealing with the linear structural break tests.

- Copula independence test is well interpreted as a correlogram equivalent when similarly applied to time series components. Different from the correlogram, the dependogram (visual representation of copula independence test) does not distinguish between the effects on AR and MA components. Nevertheless, the inference is common as the above example have shown.

- Empirical validation of the testing procedure was done on US GDP quarterly growth rate series. Compared to Andrews-Zivot test results bringing structural shift years as 1964, 1970 and 1977, copula structural break test enabled to detect the structural change taking place after the Iranian Revolution and another oil price spike in 1981. This is considered to be the evidence of the proposed test efficiency as conventional approaches did not result in detecting this recession (as the next one was only in 1990-1991).

**Annex 1. Copula-Based Time Series Independence Test.**

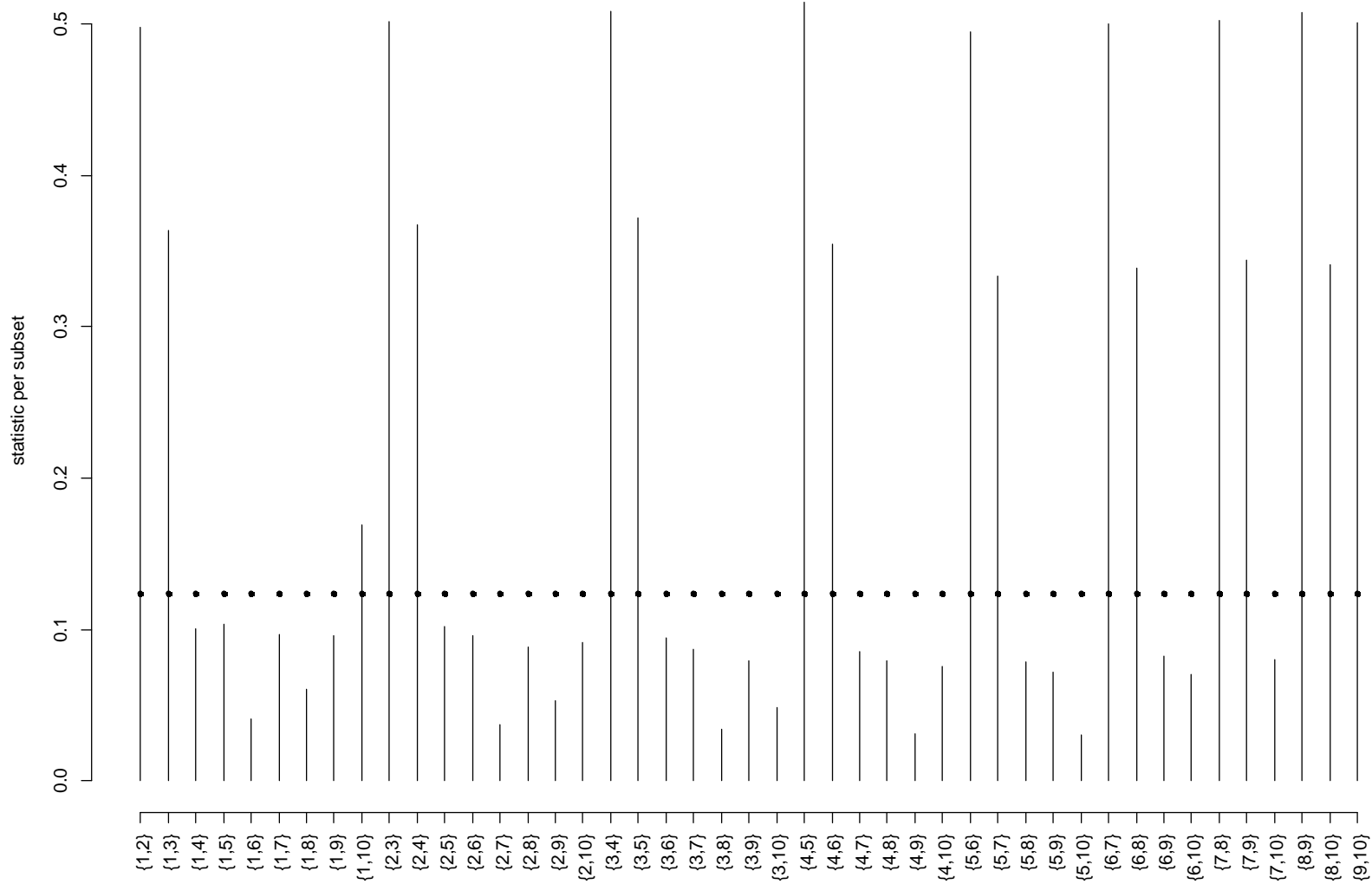

Note: 1 stands for zero lag; 2 – for the first etc.; i.e. {2;10} case presents test results for first lag and ninth lag.

**Annex 2. Andrews-Zivot Structural Break Test Output.**

### (1) Intercept

Dependent Variable: Y
Method: Least Squares
Sample(adjusted): 3 262
Included observations: 260 after adjusting endpoints

| Variable | Coefficient | Std. Error | t-Statistic | Prob. |
|---|---|---|---|---|
| C | 0.010701 | 0.001616 | 6.620455 | 0.0000 |
| DU72 | 0.008176 | 0.002253 | 3.629600 | 0.0003 |
| TR | -5.48E-05 | 1.35E-05 | -4.055784 | 0.0001 |
| Y(-1) | 0.413326 | 0.056549 | 7.309130 | 0.0000 |

| | | | |
|---|---|---|---|
| R-squared | 0.295631 | Mean dependent var | 0.016243 |
| Adjusted R-squared | 0.287376 | S.D. dependent var | 0.011544 |
| S.E. of regression | 0.009745 | Akaike info criterion | -6.408867 |
| Sum squared resid | 0.024311 | Schwarz criterion | -6.354087 |
| Log likelihood | 837.1527 | F-statistic | 35.81521 |
| Durbin-Watson stat | 2.063759 | Prob(F-statistic) | 0.000000 |

### (2) Trend

Dependent Variable: Y
Method: Least Squares
Sample(adjusted): 3 262
Included observations: 260 after adjusting endpoints

| Variable | Coefficient | Std. Error | t-Statistic | Prob. |
|---|---|---|---|---|
| C | 0.007629 | 0.001817 | 4.198986 | 0.0000 |
| TR | 4.88E-05 | 2.04E-05 | 2.396557 | 0.0173 |
| DT123 | -0.000120 | 3.46E-05 | -3.467428 | 0.0006 |
| Y(-1) | 0.413901 | 0.056890 | 7.275465 | 0.0000 |

| | | | |
|---|---|---|---|
| R-squared | 0.292606 | Mean dependent var | 0.016243 |
| Adjusted R-squared | 0.284316 | S.D. dependent var | 0.011544 |
| S.E. of regression | 0.009766 | Akaike info criterion | -6.404582 |
| Sum squared resid | 0.024415 | Schwarz criterion | -6.349802 |
| Log likelihood | 836.5956 | F-statistic | 35.29720 |
| Durbin-Watson stat | 2.071036 | Prob(F-statistic) | 0.000000 |

### (3) Intercept + Trend

Dependent Variable: Y
Method: Least Squares
Sample(adjusted): 3 262
Included observations: 260 after adjusting endpoints

| Variable | Coefficient | Std. Error | t-Statistic | Prob. |
|---|---|---|---|---|
| C | 0.009690 | 0.002202 | 4.400010 | 0.0000 |
| DU96 | 0.006899 | 0.002558 | 2.697157 | 0.0075 |
| TR | -3.63E-06 | 3.69E-05 | -0.098266 | 0.9218 |
| DT96 | -7.27E-05 | 4.08E-05 | -1.780956 | 0.0761 |
| Y(-1) | 0.399352 | 0.056946 | 7.012771 | 0.0000 |

| | | | |
|---|---|---|---|
| R-squared | 0.303651 | Mean dependent var | 0.016243 |
| Adjusted R-squared | 0.292728 | S.D. dependent var | 0.011544 |
| S.E. of regression | 0.009708 | Akaike info criterion | -6.412627 |
| Sum squared resid | 0.024034 | Schwarz criterion | -6.344153 |
| Log likelihood | 838.6415 | F-statistic | 27.79898 |
| Durbin-Watson stat | 2.047217 | Prob(F-statistic) | 0.000000 |